**Advancements in Tuning Thermoelectric Properties: Insights from Hybrid Functional Studies, Strain Engineering, and Machine Learning Models**

*Vipin K E[a,*], and Prahallad Padhan[a,b],*

[a]*Department of Physics, Nanoscale Physics Laboratory, Indian Institute of Technology Madras, Chennai 600036, India*

[b]*Functional Oxides Research Group, Indian Institute of Technology Madras, Chennai 600036, India*

\* Corresponding author at: Department of Physics, Indian Institute of Technology Madras, Chennai 600036, India

Email addresses: ph18d200smail.iitm.acin

ABSTRACT

Thermoelectric properties in topological insulator $Bi_2Se_3$ are explored with multifaceted strategies, i.e., hybrid functional with strain and artificial intelligence methodology. The assessment with the experimental band gap values recognises the limitations of conventional functional and the effectiveness of screened hybrid functionals. A thorough investigation into the impact of biaxial and uniaxial strain on thermoelectric parameters uncovers distinctive behaviours in n-type and p-type $Bi_2Se_3$, providing insights into optimal strain conditions for improved performance. Furthermore, the studies on the role of topologically non-trivial surface states (TNSS) in thermoelectric properties reveal that TNSS significantly dominate electronic transport. Dual scattering time approximation elucidates the segregation of thermoelectric transport contributions from bulk and surface states, highlighting the importance of controlling the relaxation time ratio for enhanced thermoelectric performance. Additionally, the prediction of thermoelectric properties using Random Forest and Neural Networks models showcase impressive agreement with density functional theory predictions across varying temperatures, offering a powerful tool for understanding complex temperature-dependent trends in thermoelectric properties. In summary, this interdisciplinary study presents a unique approach to advancing the understanding and optimization of thermoelectric properties in $Bi_2Se_3$. It provides a comprehensive framework for tailoring material behaviour for diverse thermoelectric applications.

1. **Introduction**

Thermoelectricity is a fascinating phenomenon where a temperature gradient within a material generates an electric voltage. One of the most significant advantages of thermoelectricity over traditional methods is its efficiency in harnessing waste heat. Traditional power generation systems lose a substantial amount of energy as heat, which is often dissipated into the environment. However, thermoelectric generators can capture this wasted heat and convert it into electricity, offering an elegant and eco-friendly solution for improving energy efficiency. This capability makes them invaluable in various sectors, including automotive, where they can enhance fuel efficiency, and industrial processes, where they can recover heat from machinery and processes. Thus, over several decades, persistent research endeavours have yielded numerous innovative strategies and theoretical frameworks to augment the performance of thermoelectric (TE) materials. These approaches encompass a spectrum of techniques, including alloying, nano-engineering, band convergence, energy filtering, hierarchical architecturing, chemical doping, and functionalization. [1] Additionally, the discipline has witnessed the introduction of methods such as strain engineering, [2] defect engineering, [3] and the creation of heterostructures. [4]

Achieving optimal thermoelectric performance is characterized by a high thermoelectric figure of merit (ZT). The ZT is calculated using the Seebeck coefficient (S), electrical conductivity($\sigma$), thermal conductivity($\kappa$), and absolute temperature(T) of a TE material. To attain a high ZT value, a delicate balance between S, $\sigma$ and $\kappa$ is required. These parameters, unfortunately, are inherently interdependent and correlated, making their independent tuning nearly impossible. Enhancing the thermopower often comes at the expense of electrical conductivity, while reducing thermal conductivity can inadvertently hinder electrical conductivity. This intricate trade-off necessitates innovative materials design and engineering approaches to break through these limitations and unlock the full potential of thermoelectric materials.

The non-trivial surface states present in topological insulators (TIs) follow time-reversal symmetry, which differentiates TIs from conventional metals and insulators. Notably, numerous materials with topologically non-trivial properties, including but not limited to $Bi_2Te_3$, $Bi_2Se_3$, and $Sb_2Te_3$, have garnered attention due to their exceptional thermoelectric performance. The thermoelectric properties of these TIs are proficient because of the heavy constituent elements, a narrow band gap, and large spin-orbit coupling. [5] The surface states of the TIs are characterized by linearly scattered Dirac cones, which contribute to the exceptionally high surface mobility. This, in turn, creates intriguing electrical transport characteristics within these materials, marked by robust carrier mobility, rendering these TIs as the upcoming candidates for implementation in thermoelectric applications. [6] Additionally, materials with heavy constituent elements manifest a propensity for soft vibrational (phonon) modes, reducing lattice thermal conductivity. This attribute assumes paramount importance in pursuing superior thermoelectric performance of these TIs. [7]

It is postulated that through the introduction of controlled levels of non-magnetic disorder or defects and the precise manipulation of the Fermi level position, the TE characteristics can be significantly tailored and optimized. [8, 9] The backscattering caused by the non-magnetic disorders and defects in TIs does not change the electronic transport, but it influences phonon transport strongly. Thus, a successful decoupling of these two modes of transport is achieved, thereby realizing the "phonon-glass, electron-crystal" paradigm, ultimately leading to heightened TE performance in TIs. [9, 10] In a study, Yong et al. observed that the lifetime of surface states in TIs exhibits a strong dependence on energy levels, resulting in notable and unconventional Seebeck effects with a polarity opposite to the Hall effect. [11] For $Bi_2Se_3$ films with a thickness exceeding six quintuple layers, the relaxation time of the topological surface states within the bulk energy gap is on the order of hundreds of femtoseconds, marking a substantial increase compared to bulk states. [12] This high surface relaxation time, stemming from the robust nature of topological surface states, leads to the attainment of maximal ZT values, which can be further optimized to approximately 2.0 under realistic physical conditions for 3QL (quintuple layer) $Bi_2Te_3$. [9] Another study by Peter et al. reported that the optimization of p-doped $Bi_2Se_3$ offers significant advantages over $Bi_2Te_3$ concerning thermoelectric properties within the temperature range of 400 to 600 K. [13] Numerous strategies have been employed to enhance the thermoelectric performance of $Bi_2Se_3$, including strain engineering, nanostructuring and doping. Doping with elements such as copper, antimony, and nickel has increased ZT values of 0.54, 0.036, and 0.57, respectively. [14, 15, 16] Additionally, nanostructuring modifications, such as nanoplates and nanosheets of $Bi_2Se_3$, have yielded improved performance with ZT values of 0.14 and 0.17, respectively. [17, 18]

Using a combination of van der Waals Density Functional Theory (DFT) and semi-classical Boltzmann theory, the research has revealed intriguing insights into the ZT for a single QL of $Bi_2Se_3$. The ZT was found to be 0.27, surpassing the bulk value of 0.10, and further enhancement to 0.30 was achieved by introducing a 2.5 % compressive strain. [19] In the context of n-doped $Bi_2Se_3$ and n-doped $Bi_2Te_3$, it was observed that the in-plane power factor could be increased by applying compressive uniaxial strain. [10] Strain-dependent investigations on $TiS_2$ indicated a figure of merit of up to 0.95 (0.82) under an 8 % tensile strain for p-type (n-type) doping. Additionally, these studies revealed a notable indirect−direct−indirect band gap transition with increasing tensile strain. [20]

Further exploration extended to two-dimensional transition metal dichalcogenide monolayer WS$_2$, wherein a 77 % increase in the power factor was observed for n-type material upon applying uniaxial compressive strain. Moreover, a decrease in lattice thermal conductivity with rising temperature led to an almost 40 % increase in the ZT product under applied uniaxial compressive strain. [21] Similarly, for InSe monolayers, a 6 % applied strain resulted in a substantial decrease in lattice thermal conductivity from 25.9 to 13.1 WK$^{-1}$m$^{-1}$. [22] The investigation of strain-induced thermoelectric performance of PbTe unveiled the potential for optimal ZT values of 3.88 for p-type material under 0.5 % compressive strain and 3.05 for n-type material under 2.0 % compressive strain. [23] These findings underscore the remarkable versatility and promise of strain engineering in tailoring and enhancing the thermoelectric properties of a wide range of materials.

Recent research highlights the applicability of the Generalized Gradient Approximation (GGA) with van der Waals (vdW) treatment as the most suitable method for structural optimization in the case of Bi$_2$Se$_3$ and Bi$_2$Te$_3$. While there is a slight inclination towards the meta-GGA approach in terms of accuracy, the preference leans towards GGA due to the associated computational advantages in capturing the behavior of TIs. [24] There has been a growing interest in employing hybrid functionals that combine local density approximation (LDA) or GGA with the Hartree-Fock approximation (HFA) to overcome the limitations inherent in semilocal approximations. [25] Hybrid-DFT computations have successfully accurately estimated band gaps and characterized defects in semiconductors, insulators, and their interfaces, all while mitigating the computational costs associated with GW calculations. [26] A notable study employing the B3PW91 hybrid functional for 27 related binary and ternary semiconductors reported band gap predictions with a mean average deviation (MAD) of merely 0.09 eV—a substantial improvement compared to the results obtained with LDA and GGA density functionals. [27] Furthermore, investigations have shown that the HSE (Heyd-Scuseria-Ernzerhof) functional can enhance its accuracy for materials with significant band gaps without adversely affecting its performance for semiconductors, accomplished by incorporating a more significant portion of Hartree-Fock exchange in its short-range treatment. [28] Notably, HSE functional studies focused on small-band-gap materials have yielded band-gap predictions closely aligned with experimental values, boasting a mean relative error (MRE) of just 0.7%, thus establishing a degree of comparability with, or even superiority to, the outcomes achieved through the GW approximation. [29] A multitude of theoretical investigations and experimental studies are actively pursued to elucidate the precise band structure of $Bi_2Se_3$, with a consensus emerging that it is a narrow band gap material whose properties are strongly influenced by intrinsic spin-orbit coupling. Reported values for the band gap of $Bi_2Se_3$ have exhibited a range spanning from 1.6 eV to 3.6 eV. [30, 31, 32] Nevertheless, the nature of the band gap and the material's underlying structure remain subjects of ongoing debate.

Contemporary research explores the machine learning (ML) approach as a promising alternative to DFT for material properties predictions. ML techniques offer accurate results with significantly reduced computational time, unveiling novel correlations among material descriptors. This accelerates the discovery and development of materials, particularly in thermoelectric (TE) research. [33, 34] Supervised learning, a prevalent ML method in materials science, leverages relevant data with known target properties. This approach has been extensively applied in thermoelectric material development, predicting key properties such as Seebeck coefficient (S), [35] power factor ($S^2\sigma$), [36] lattice thermal conductivity ($\kappa_L$), [37] and ZT values. [38]

The present study addresses the band gap quandary of $Bi_2Se_3$ within the framework of DFT, employing various exchange-correlation functionals, with a particular emphasis on hybrid functionals. Further, the incorporation of strain with the hybrid functionals modifies the band structure of $Bi_2Se_3$, indicating the strain is a potent tool for controlling different features that could be valuable for the development of new capabilities and creative technologies. The unstrained and strained thermoelectric properties of $Bi_2Se_3$ obtained from the DFT are close to the value predicted using the Random Forest (RF) and Neural Network (NN) models, which demonstrate their efficacy in enhancing efficiency and reducing computational costs.

2. **Theoretical Methodology**

**Density functional theory**

The hexagonal crystal structure of $Bi_2Se_3$ was created with VESTA [39] software using the lattice parameters of the nanocrystals synthesized using the hot injection method. [40] The band structure calculations of $Bi_2Se_3$ were performed using the Quantum Espresso (QE) simulation package. [41] The GGA approximations are made with PBE exchange-correlation functional using scalar relativistic and fully relativistic projector-augmented-wave-type pseudopotential. For SCAN and Hybrid functionals, the band structure calculations were performed using the norm-conserving scalar relativistic and fully relativistic pseudopotential. The plane wave cut-off and the charge density cut-off were set to be 50 Ry and 200 Ry for the plane wave basis set. The dense Monkhorst−Pack k-mesh grid of 15 × 15 × 3 was used for PBE and SCAN functionals, and 8 × 8 × 2 was used for hybrid calculations with q mesh sampling for the Hartree-fock operator of 4 × 4 × 2. The structure undergoes relaxed calculations to obtain the atomic positions near their equilibrium values with force on each atom of <10$^{-3}$ eV/Å. The wannier90 software was used to project the p orbitals in bulk $Bi_2Se_3$ for the corresponding band structure.

**Boltzmann transport theory**

The phonon dispersion and thermal properties calculations of the $Bi_2Se_3$ were performed using the PHONOPY [42] package interfaced with QE. [41] The PHONOPY was used to calculate phonons at harmonic and quasi-harmonic levels with small atomic displacements at the constant volume of the supercell. The atomic vibrations are solved with the second-order terms in the crystal potential energy as the harmonic approximation. The higher-order terms are treated by perturbation theory. A collection of force sets is created from the change in potential energy, in which crystal symmetry is utilized to improve the force constants' numerical accuracy and reduce the computational cost; phonon dispersion is calculated from the collected force sets. The lattice thermal conductivity was calculated using the PHONO3PY [43] software package interfaced with the QE package. The Q-point sampling meshes of $19 \times 19 \times 19$ were used in the thermal conductivity calculations, and $32 \times 32 \times 8$ was used for phonon dispersion.

**Machine learning schemes**

The dataset presents a meticulous curation of a vast thermoelectric materials dataset from experimental observations collected from databases and references, spanning 5224 distinct materials across 67 elements, tailored for machine learning applications. Utilizing Matminer and Pymatgen libraries, [44, 45] a comprehensive feature generation process captures diverse material characteristics. Temperature was added as an additional feature for the material composition. A refined correlation matrix reduces the feature count, enhancing dataset efficiency. The dataset undergoes a shuffle for unbiased model training, followed by a 90:10 train-test split. Feature normalization ensures fair comparisons across scales, laying the groundwork for robust machine learning model development. NN [46] architecture was implemented using TensorFlow [47] to predict material properties, specifically, Electrical

Conductivity, Seebeck Coefficient, power factor, and ZT. The NN model, featuring varied layers and neurons, employed mean absolute error (MAE) as the loss function. Early stopping criteria were included during the 30,000-epoch training process to prevent overfitting. Evaluation, based on $R^2$ values, showcased the NN's high accuracy in predicting thermoelectric properties. The RF [48] model, equipped with meticulously tuned hyperparameters, was used for our study. Evaluation of the model was based on $R^2$ values.

3. **Results and discussion**

Thermoelectric properties of $Bi_2Se_3$ are simulated using the 3QLs unit cell and lattice parameters. A schematic of a bc-plane of the 2QLs of hexagonal $Bi_2Se_3$ is shown in Figure 1a, where a unit of five layers arranged in the sequence of $Se1 - Bi - Se2 - Bi - Se1$ forms QL. Since the simulated band structure of $Bi_2Se_3$ depends on the exchange-correlation functionals, the band gap ($E_g$) of the same $Bi_2Se_3$ unit cell was determined for various exchange-correlation functionals. The most widely used Perdew−Burke−Ernzerhof (PBE) parametrized exchange-correlation functional, [49] shows the $Bi_2Se_3$ direct band gap of 0.2 eV and the band inversion at the Γ-point with the inclusion of spin-orbit coupling (SOC). [50] However, the PBE is not a universal improvement over the local density approximation (LDA) as, in both cases, their accuracy relies on some cancellation of error in exchange and correlation, which is system-dependent. [51] As compared to the LDA, the generalized gradient approximation (GGA) provides better results, but the binding energy is underestimated. [51] The residual self-interaction is one of the most significant causes of the underestimation of the $E_g$ in LDA and GGA-based DFT calculations, which results from the electron-electron interaction in the Hamiltonian. The self-interaction energy is positive, and it raises the energy of localized states and results in the delocalization of electrons.

The accuracy over the GGA functional is further improved in the meta-GGA functional, which includes higher power up to the fourth order in the enhancement factor. The meta-GGA's are much more successful in predicting material properties than LDA and GGA. The strongly constrained and appropriately normed (SCAN) semi-local density functional is a nonempirical semi-local functional that satisfies all known possible exact constraints and is appropriately normed on systems for which semi-local functionals can be exact or extremely accurate. [52] Since SCAN functional strongly depends on the Laplacian of the density and on kinetic energy densities, which makes them sensitive to the degree of localization of electrons and helps to distinguish covalent bonds in $Bi_2Se_3$. However, when the exchange-correlation hole is considered, the electrons are shared between the covalent bonds rather than the density being localized near the reference electron, which makes meta-GGA less predictive for $Bi_2Se_3$ crystal. In addition, the semi-local density functionals over-delocalize the electrons, making it difficult for the localized subsystems like the defect, surface states and d and f-block elements. [53]

In addition to the GGA and meta-GGA, we have also used the hybrid functionals, which give very physically optimized effective-correlation potentials and predict more accurate material behaviour. [54] The hybrid functionals are the mixture of exact exchange of the Hartree–Fock (HF) and DFT exchange-correlation term. In the Hartree–Fock method, the exact exchange energy depends on the individual particle states, and this exchange cancels the self-interaction term but does not take into account the electronic correlation. The exchange itself is long-ranged, decaying only as $r^{-1}$, and not screened, leading to high excitation energies and a large overestimation of the band gap. The use of hybrid functional is a pragmatic approach since both methods give errors of the opposite sign compared to the experimental data. The hybrid exchange-correlation functionals improved many molecular properties relative to semi-local functionals and are widely used in computational chemistry. [55] The band structure calculations were performed using the B3LYP (Becke, 3-parameter, Lee–Yang–Parr), [56] PBE0, [49] GAUPBE (GaussianPBE) [57] and HSE (Heyd–Scuseria–Ernzerhof) [58] hybrid functionals.

For 0.1 percentage of the Hatree-Fock exchange in different hybrid functionals, the $E_g$ of the $Bi_2Se_3$ varies between ~ 0.3 eV to ~ 0.5 eV (Figure 1b). As the percentage of the Hatree-Fock exchange in the hybrid functional increases, the $E_g$ of the $Bi_2Se_3$ increases. The incorporation of the exact exchange allows increased localization of the electronic states and tuning of the band gap. However, the nonlocal exchange interaction has an unphysical and extremely slow spatial decay, and the asymptotic decay of the exchange potential for atomic and molecular systems is incorrect. The exact exchange potential decays asymptotically as $r^{-1}$, [59] while that of a hybrid functional with a fraction 's' of nonlocal exchange decays as (s $r^{-1}$). This incorrect long-range decay is believed to be responsible for errors in describing various material properties, especially in B3LYP and PBE0.

The long-range part of the HF exchange is minimized by introducing a screened coulomb potential. Thus, the coulomb operator is split into long-range (LR) and short-range (SR) components, where, the short-range exchange component is not screened. The HSE is one of the popular screened hybrid functional, which contains the screened Coulomb potential hybrid density functional of the form;

$$E_{XC}^{HSE} = aE_X^{HF,SR}(r_{sp}) + (1-a)E_X^{PBE,SR}(r_{sp}) + E_X^{PBE,LR}(r_{sp}) + E_C(r_{sp}), (1)$$

where '$r_{sp}$' is an adjustable parameter, usually called the range separation parameter [58] and 'a' is the percentage of HF exchange, 'Ex' is the exchange and 'Ec' is the correlation energy, respectively. The $E_{XC}^{HSE}$ is equivalent to $E_{XC}^{PBE0}$ for $r_{sp}$ = 0 and asymptotically reaches $E_{XC}^{PBE}$ for $r_{sp} \to \infty$. GAUPBE is another kind of screened HF that uses the Gaussian-attenuating exact exchange in the coulomb potential.

The $r_{sp}$ in the HSE is generally system-dependent; [60] thus, the suitable value $r_{sp}$ of the $Bi_2Se_3$ was determined by calculating the band structure of the unit cell for various $r_{sp}$. For a = 0.25 and $r_{sp}$ = 0.11 a.u., the $E_g$ of the $Bi_2Se_3$ is ~ 0.51 eV, which decreases with the increase in the $r_{sp}$ (Figure 1c). The variation of $E_g$ with $r_{sp}$ indicates that the $r_{sp}$ in the range of 0.2 a.u. to 0.3 a.u. is a good choice for experimentally observed $E_g$ of the $Bi_2Se_3$. For a = 0.1 and $r_{sp}$ = 0.25, the $E_g$ of the $Bi_2Se_3$ is 0.31 eV, which is close to the band gap observed in the experiment. [61] The comparison of $E_g$ of the $Bi_2Se_3$ observed from the experiments strongly indicates that the short-range screened hybrid can describe the band gap of $Bi_2Se_3$. Thus, one does not need to resort to more computationally expensive global hybrids functional like B3LYP and PBE0.

The bulk band topology has been shown to be strain sensitive, under uniaxial strain, the shift from the topological to the trivial phase was projected to occur for topological bulk materials [62]. The strain has also been proposed as an effective parameter to drive a topological phase transition (TPT) in the topological insulators (Tis). [10] The strain-dependent band structure of the $Bi_2Se_3$ is studied by adopting the HSE hybrid functional. The biaxial and uniaxial strain was induced by varying the in-plane (a = 4.143 Å) and out-of-plane (c = 28.636 Å) lattice parameters, respectively, up to ± 6 %, where +ve and -ve signs represent tensile and compressive strain, respectively.

For uniaxial compressive strain and strain-free state, the band structures show an indirect band gap (Figure 2a, b, c), which becomes direct under the uniaxial tensile strain (Figure 2d, g). Interestingly, under the 6 % uniaxial tensile strain, the band structure of $Bi_2Se_3$ predicts a topological phase transition (closing of the band gap) (Figure 2g), which, interestingly, is approximately consistent with the prediction by Young et al. [63] In contrast, the band gap under the biaxial strain remains indirect (Figure 3a-e). As either the uniaxial or biaxial strain varies from -6 % to 6 %, the band gap decreases (Figure 4). In Figure 2g, the Dirac cone typical of the phase transition is clearly visible at 6%, the transition point. Such strains are, of course, difficult to achieve experimentally. However, a recent experimental study demonstrates the occurrence of 9.6 % strain along 'a' and 19.6 % along 'c' through the intercalation of 60 % zero-valent Cu into $Bi_2Se_3$ nanoribbons without disrupting the host lattice [64].

The thermoelectric properties of the $Bi_2Se_3$ were calculated with the dual scattering time approximation, and short-range screened hybrid functional. In addition, the contributions from the bulk and surface to the total thermoelectric properties of the $Bi_2Se_3$ are separated. The thermoelectric properties have been tuned and optimized by controlling the position of the Fermi level ($E_F$), which in turn can be achieved by introducing defects and dopants into the system. [65] The thermoelectric properties and electronic transport properties of the bulk $Bi_2Se_3$ can be determined within the framework of Boltzmann transport in the linear response regime and diffusive limit of transport, respectively, once the band structure is calculated. The band structure of the bulk $Bi_2Se_3$ was calculated to determine the transport distribution function (TDF), which is given by [66]

$$\Xi_{ij}(\varepsilon) = \frac{1}{V} \sum_{nk} v_i(n,k) v_j(n,k) \tau_{nk} \delta(E - E_{n,k}), (2)$$

where the summation is over all bands 'n' and over all the Brillouin zone (BZ), $E_{n,k}$ is the energy for the band 'n' at 'k', and $v_i$ is the i$^{th}$ component of the band velocity at (n, k) and $\tau_{n,k}$ is the constant relaxation time. The in-plane Seebeck coefficient 'S', electrical conductivity

'σ', and electronic thermal conductivity '$\kappa_e$' are then expressed as integrations of TDF as follows [66]

$$[\sigma]_{ij}(\mu,T) = e^2 \int_{-\infty}^{+\infty} \left( \frac{-\partial f(E,\mu,T)}{\partial E} \right) \Xi(E) dE \quad (3)$$

$$[\sigma S]_{ij}(\mu,T) = \frac{e}{T} \int_{-\infty}^{+\infty} \left( \frac{-\partial f(E,\mu,T)}{\partial E} \right) (E-\mu) \Xi(E) dE \quad (4)$$

$$[\kappa_e]_{ij}(\mu,T) = \frac{1}{T} \int_{-\infty}^{+\infty} \left( \frac{-\partial f(E,\mu,T)}{\partial E} \right) (E-\mu)^2 \Xi(E) dE \quad (5)$$

Here, i and j are cartesian indices, [σS] denotes the matrix product of the two tensors, and ∂f/∂E is the derivative of the Fermi–Dirac distribution function with respect to the energy. [67]

The strain-dependent electrical conductivity σ(ϵ) was calculated using Eq. 3 for n-type and p-type carriers of the $Bi_2Se_3$ at 300 K. The σ of the n-type carrier of the $Bi_2Se_3$ under -6 % uniaxial strain is $1.69 \times 10^4$ S/m. The σ increases linearly as the ϵ decreases from -6 % to zero and increases to 6 %. In contrast, the σ of the p-type carrier of the $Bi_2Se_3$ under -6 % uniaxial strain is $12.83 \times 10^4$ S/m, decreases as the ϵ decreases to zero, and then gradually increases with the increase in the ϵ up to 6 % (Figure 5a). The S of the n-type carrier of the $Bi_2Se_3$ under -6 % uniaxial strain is -176 × μW $K^{-1}$, which decreases as the strain decreases to zero, becomes -35 μW $K^{-1}$ for 2 % strain and then remains constant up to 6 %. The S of the p-type carrier of the $Bi_2Se_3$ under -6 % uniaxial strain is S 86 μW $K^{-1}$, which increases as the strain decreases to zero and then decreases with the increase of strain to 6 % (Figure 5b). Interestingly, the power factor ($S^2σ$) of the n-type carrier of $Bi_2Se_3$ decreases as the uniaxial strain varies from -6 % to 6 %. The $S^2σ$ of the p-type carrier is larger than the n-type carrier of the $Bi_2Se_3$. The $S^2\sigma$ of the p-type carrier is almost constant under compressive strain, while it decreases with the increase in the tensile strain (Figure 5c).

Under the compressive biaxial strain, the σ of the n-type carrier of the $Bi_2Se_3$ is relatively higher compared to the uniaxial strain. Under -6 % strain, the σ is ~ 6.34 × $10^4$ S/m, which reduces to 2.16 × $10^4$ S/m for the zero strain state. On increasing the biaxial strain to 2 %, the σ further reduces to 0.74 × $10^4$ S/m and remains the same for the higher strain. The σ under the tensile biaxial strain is smaller than the σ under the uniaxial tensile strain. On the other hand, the σ of the p-type carrier of the $Bi_2Se_3$ is almost independent of the strain between - 6 % to 5 %. However, at 6 % biaxial strain, the σ of the p-type carrier of the $Bi_2Se_3$ is increased to ~ 4.31 × $10^4$ S/m (Figure 5e). The S of the n-type carrier of the $Bi_2Se_3$ varies between - 80 μW $K^{-1}$ to -150 μW $K^{-1}$. Similarly, the S of the p-type carrier of the $Bi_2Se_3$ changes from 118 μW $K^{-1}$ to 200 μW $K^{-1}$. For the n-type carrier of the $Bi_2Se_3$, the S is higher under the biaxial compressive strain; in contrast, the higher S is observed under the tensile strain for the p-type carrier of the $Bi_2Se_3$ (Figure 5f). As the biaxial strain varies from -6 % to 6 %, the $S^2\sigma$ of the n-type carrier of $Bi_2Se_3$ decreases, similar to the uniaxial strain. The $S^2\sigma$ of the n-type carrier of $Bi_2Se_3$ under the biaxial strain is relatively larger than the uniaxial strain. On the other hand, the $S^2\sigma$ of the p-type carrier of the $Bi_2Se_3$ is maximum for the un-strain state and decreases either with compressive or tensile strain. The S and $S^2\sigma$ of the p-type carrier of $Bi_2Se_3$ are larger than the n-type carrier of $Bi_2Se_3$, irrespective of the nature of the strain (Figure 5g).

The thermoelectric materials generate electricity once exposed to the temperature gradient. The performance of the thermoelectric conversion of the material is determined by the efficiency, i.e., the dimensionless figure of merit $ZT = \frac{\sigma S^2 T}{\kappa}$. The strain-dependent σ and S were calculated by adopting HSE hybrid functional. However, the lattice thermal conductivity κ of only strain-free $Bi_2Se_3$ structure was calculated by using the PBE functional due to the high computational time of the HSE hybrid functional. At 300 K, the ZT (~0.15) of the p-type carrier is larger than the n-type carrier of $Bi_2Se_3$, in addition, the ZT of the strain-free p-type carrier of $Bi_2Se_3$ is maximum (Figure 5d,h) irrespective of uniaxial or biaxial strain. The ZT for the p-type carrier of $Bi_2Se_3$ decreases after the application of either compressive or tensile strain. However, the ZT of the n-type carrier of $Bi_2Se_3$ increases for the compressive strain but decreases for the tensile strain compared to the strain-free state ZT.

As seen in Figure 2, the band structure of $Bi_2Se_3$ strongly depends on the applied strain. The presence of an indirect band gap, direct band gap, and topologically non-trivial surface states (TNSS) are expected to play a vital role in determining the transport properties of the $Bi_2Se_3$, i.e., TDF. The TDF of $Bi_2Se_3$ depends on the state's relaxation time (τ) (Eq. 2). In addition, the surface states are protected from backscattering. Thus, the $Bi_2Se_3$ have different conducting channels associated with the surface states and bulk states. The band structure of the n-type 3-QLs of $Bi_2Se_3$ was calculated using the slab model (Figure 6). Inside the bulk band gap, the surface state relaxation time ($τ_S$) is more than the bulk state relaxation time ($τ_B$), normally located outside the bulk band gap. The $τ_B$ and $τ_S$ strongly depend on the band gap and the nature of the states dispersion. In $Bi_2Se_3$, the in-plane $τ_B$ is 2.7 fs, [10] which is smaller compared to the $τ_S$ (2.2 fs). [68] The $τ_S$ of $Bi_2Te_3$ calculated using the experimental mean free path, and Fermi velocity is 550 fs, [9] which is also larger than its $τ_B$ is 22 fs. [10] A variation in the scattering rate for the bulk and surface states has been observed in the Terahertz spectroscopy measurements. [68] Thus, the conduction properties in the TIs can be expressed in terms of the relaxation time ratio $τ = \dfrac{τ_S}{τ_B}$. The τ can be controlled by establishing distortion of TNSS, which can be achieved by introducing strain, i.e., by substituting impurities or creating disorder. The τ may be realized by random alloying, [69] stacking disorder, [70] isoelectronic substitution, [71] or non-magnetic impurities. [72]

The thermoelectric properties of n-type 3-QLs of $Bi_2Se_3$ are studied using the dual scattering time approximation. The total electrical conductance ($\sigma_t$) and Seebeck coefficient ($S_t$) are considered as the contribution from the surface and bulk, i.e., $\sigma_t = \sigma_S + \sigma_B$ and $S_t = (S_B \sigma_B + S_S \sigma_S)/(\sigma_B + \sigma_S)$, where $\sigma_S$ and $S_S$ are the surface and, $\sigma_B$ and $S_B$ are the bulk contributions. [11] The influence of the TNSS is studied as a function of τ by considering the variation of $\tau_S$ with a negligible change of $\tau_B$, i.e., $\tau_B = 2.7$ fs.

The $\sigma_s$ (Figure 7a) is two orders larger than the σ observed in Figure 5a,e. In addition, with the increase in τ, the $\sigma_S$ increases over the $\sigma_B$, and the electron transport in the $Bi_2Se_3$ slab is dominated by the surface states. Interestingly, the $S_b$ and $S_S$ are opposite signs, which provides an alternative approach for the manipulation of the thermoelectric properties of the TIs Figure 7b. As the $\tau_B$ is smaller than $\tau_S$, the band edge turns out to be a carrier filter. Moreover, the $S_s$ becomes larger than $S_b$ due to the increase in the $\sigma_s$. The electrons below and above bulk edges have significantly different contributions and are unable to cancel each other out, as shown by Eq. 3. This results in an $S_S$ with an abnormal sign (positive for n-type systems) and a high absolute value greater than 150 $\mu VK^{-1}$.

A total power factor of $2956 \times 10^{-4}$ $Wm^{-1}K^{-2}$ is achieved at the relaxation time ratio of 148, which is 2 orders higher than the bulk counterpart. A maximum ZT value of 0.45 is achieved and can be further increased by increasing the relaxation time ratio. A major factor that limits high ZT values of $Bi_2Se_3$ is the electronic contribution to thermal conductivity ($\kappa_e$). With the increase in the relaxation time ratio, both the 'σ' and '$\kappa_e$' also increase to higher values. Controlling of '$\kappa_e$' is crucial in enhancing the ZT. This high electronic part of thermal conductivity is a unique property of topological insulators and is related to the fact that the surface states in these materials are very mobile and conductive.

Thermoelectric properties of $Bi_2Se_3$ were further predicted by adopting RF and NN models at different temperatures and compared with the DFT results. The prediction accuracy of thermoelectric properties of $Bi_2Se_3$ using RF and NN models is evaluated in terms of $R^2$ values and shown in Figure 8. The high $R^2$ values confirm the usefulness of ensemble learning, highlighting its prospective for advancing predictive accuracy in the realm of materials science.

The predictions of σ of $Bi_2Se_3$ by RF and NN models closely mirror the trends observed in DFT across the plotted temperature spectrum (Figure 9a). Although the NN underestimates the σ of $Bi_2Se_3$ throughout the calculated temperature regions, the RF model predictions range (19904 s/m – 48340 s/m) falls to that of DFT estimates. Noteworthy agreement is observed at 500 K, where RF predictions closely match DFT values.

The RF and NN demonstrate commendable agreement with DFT predictions for the Seebeck coefficient of $Bi_2Se_3$ (Figure 9b). Although NN predictions display a slightly higher absolute magnitude compared to that of the DFT estimates, the overall alignment suggests a successful capture of the complex functional relationships inherent in the Seebeck coefficient. On the other hand, the RF model clearly describes the temperature-dependent Seebeck coefficient and matches closely with the DFT results. RF tends to align more closely with DFT at lower temperatures, while the NN exhibits larger deviations. Interestingly, NN predictions converge with DFT results as the temperature rises, indicating a temperature-dependent convergence pattern. This reveals the ML models' ability to accurately reproduce the temperature-dependent behaviour of this thermoelectric property.

Examining the entire temperature range, it becomes evident that the RF model predictions are generally closer to the DFT values compared to the NN predictions. For temperatures between 200 and 700 K, the RF model consistently yields power factor values that more closely align with the DFT results, indicating a better performance of the RF model in approximating the power factor for $Bi_2Se_3$ over this temperature range (Figure 9c). A strong agreement between the DFT estimates and ML predictions of ZT of the $Bi_2Se_3$ is observed (Figure 9d). Notably, at 700 K, both ML model's predictions of ZT of $Bi_2Se_3$ closely match the DFT values, showcasing their effectiveness. NN shows good predictability at low temperatures, while RF does at high temperatures. The consistent alignment across the calculated temperatures underscores the ability of ML models to effectively capture the complex interdependencies between thermopower, electrical conductivity, and thermal conductivity, which is crucial for assessing overall thermoelectric performance.

The ML models demonstrate a notable capacity to capture temperature-dependent trends in thermoelectric properties. While predictions at extreme temperatures may exhibit slight deviations, the overall performance indicates a successful adaptation to varying temperature conditions. This highlights the importance of comprehensive training datasets covering the entire temperature range of interest. Both RF and NN models exhibit comparable performance in predicting thermoelectric properties, with RF showing a marginal advantage in certain scenarios, especially at lower temperatures. The nuanced sensitivity of NN predictions may arise from the model architecture and training data. Consequently, the choice between the two models could be guided by specific application requirements, with both demonstrating efficacy in approximating complex material responses across diverse temperature conditions.

4. **Conclusion**

In conclusion, our investigation into the thermoelectric properties of topological insulator $Bi_2Se_3$ encompasses a diverse range of approaches, each shedding light on key aspects of material behaviour. Hybrid density functional studies reveal the inadequacies of traditional methods and underscore the promise of screened hybrid functionals, exemplified by HSE, in aligning with experimental band gap values. The comprehensive exploration of biaxial and uniaxial strain effects on thermoelectric parameters offers nuanced insights into the optimal strain conditions for distinct doping types. The crucial role of TNSS in electronic transport is shown through a dual scattering time approximation, emphasizing the dominance of TNSS over bulk states. The findings underscore the significance of controlling the relaxation time ratio for achieving enhanced thermoelectric performance. The presented research thus contributes to the evolving understanding of TNSS impact on thermoelectric properties, providing a basis for further exploration and refinement of topological insulators. Moreover, the integration of machine learning models, including RF and NN, into the study demonstrates their effectiveness in predicting temperature-dependent trends in thermoelectric properties. These models exhibit notable alignment with DFT predictions, showcasing their potential as valuable tools for rapid and accurate material property predictions. Collectively, this interdisciplinary investigation forms a cohesive framework that advances our knowledge of tuning thermoelectric properties in $Bi_2Se_3$. The synergy between hybrid functional studies, strain engineering insights, and machine learning models offers a comprehensive perspective, paving the way for informed material design and optimization. The presented findings hold promise for the development of next-generation thermoelectric materials with tailored properties for diverse applications.

AUTHOR INFORMATION:

Corresponding Author: ph18d200@smail.iitm.ac.in

Declaration of Competing Interest

The authors declare that they have no known competing financial interests or personal relationships that could have appeared to influence the work reported in this paper.

Data availability

The data that support the findings of this study are available from the corresponding authors upon reasonable request.

Acknowledgements:

We would like to thank the SEED faculty grant of the Indian Institute of Technology (IIT) Madras for providing experimental setup and use of the computing resources at HPCE, IIT Madras.

Figure - 1

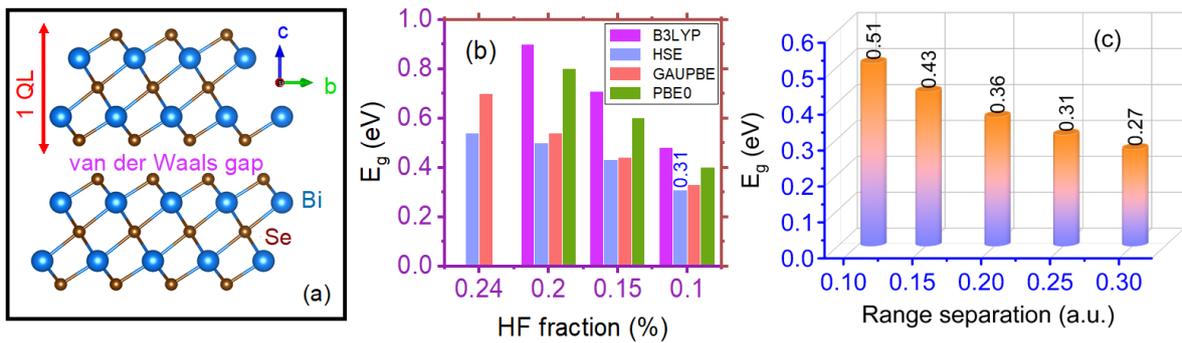

Figure 1: (a) Schematic of crystal structure of the $Bi_2Se_3$. (b) Change of band gap of the of the $Bi_2Se_3$ with the different percentage of the Hatree-Fock exchange in various hybrid functionals. (c) Variation of band gap of the $Bi_2Se_3$, with the range separation parameter in the HSE hybrid functionals.

Figure 2

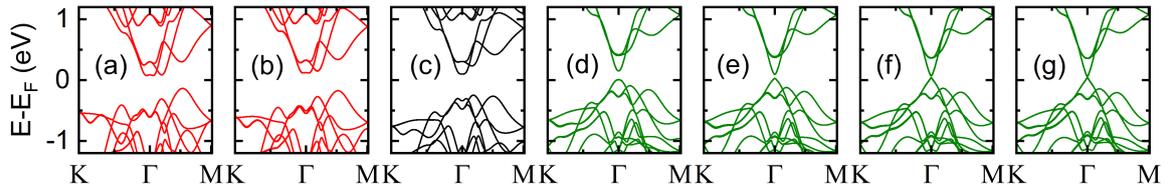

Figure 2: Band structure of Bi$_2$Se$_3$ calculated using HSE hybrid functionals for different uniaxial strain [(a) -6 %, (b) -4 %, (c) 0 %, (d) 4 %, (e) 5 %, (f) 5.5 %, and (g) 6 %].

Figure 3

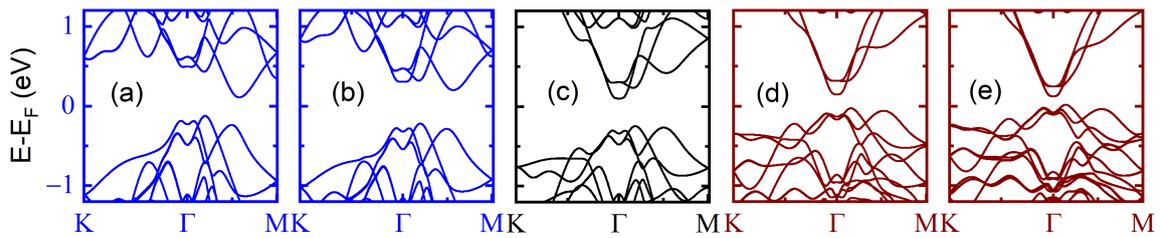

Figure 3: Band structure of Bi$_2$Se$_3$ calculated using HSE hybrid functionals for various biaxial strain [(a) -6 %, (b) -4 %, (c) 0 %, (d) 4 %, and (e) 6 %].

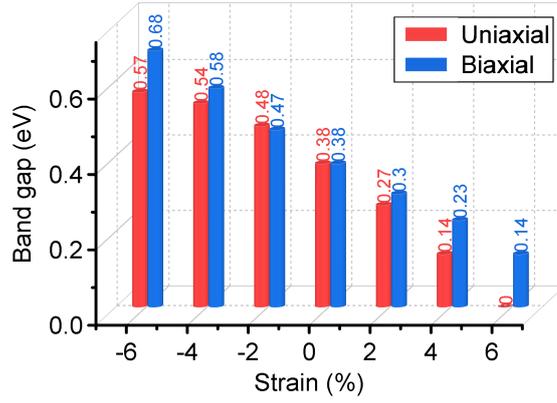

Figure 4: Variation of band gap of $Bi_2Se_3$ calculated using HSE hybrid functionals with uniaxial and biaxial strain.

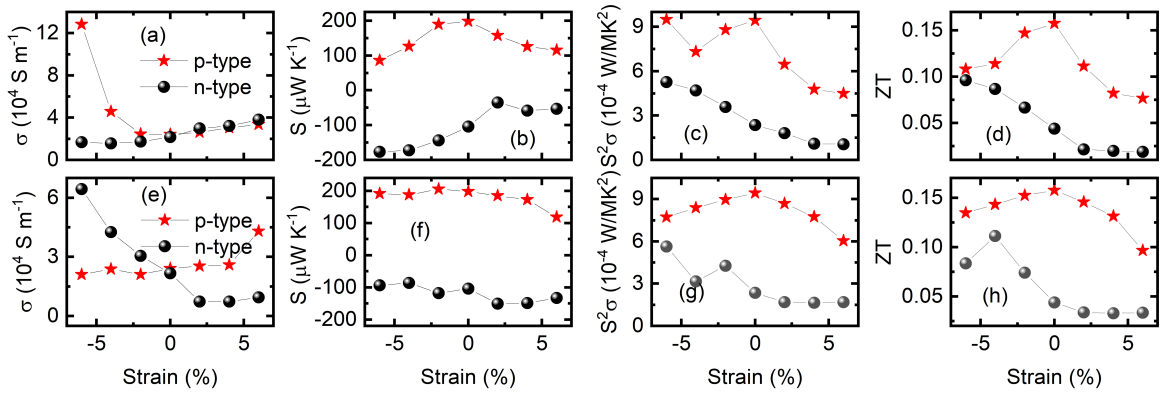

Figure 5: Electrical conductivity, Seebeck coefficient, power factor and figure of merit of the n-type and p-type $Bi_2Se_3$ for different values of uniaxial [(a), (b), (c) and (d)] and biaxial [(e), (f), (g) and (h)] strain.

Figure 6

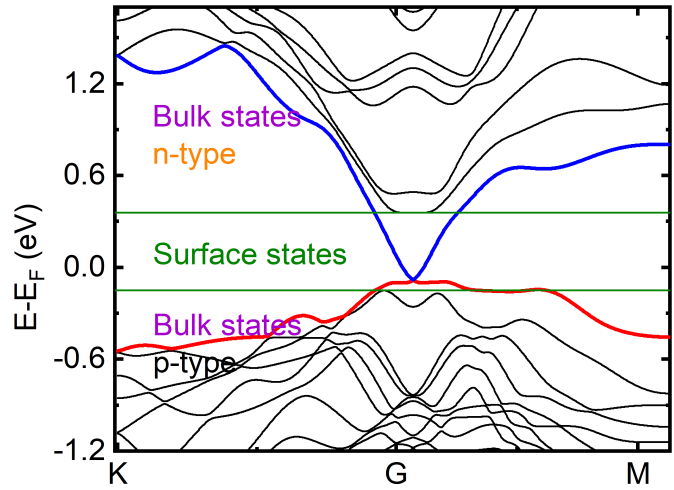

Figure 6: n-type and p-type bulk states and surface states are indicated in the surface band structure of the 3 QLs $Bi_2Se_3$ with 6 % uniaxial strain.

Figure 7

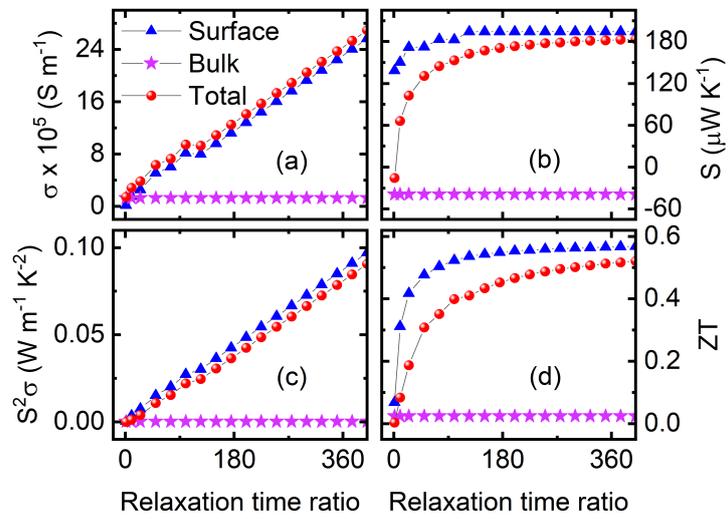

Figure 7 : Total, surface and bulk (a) electrical conductivity, (b) Seebeck coefficient, (c) power factor and (d) figure of merit of $Bi_2Se_3$.

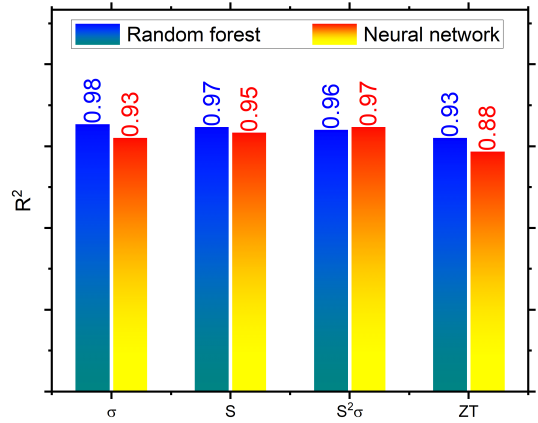

Figure 8 : Comparison of $R^2$ values found while using random forest and neural network models for prediction of electrical conductivity, Seebeck coefficient, power factor and figure of merit of $Bi_2Se_3$.

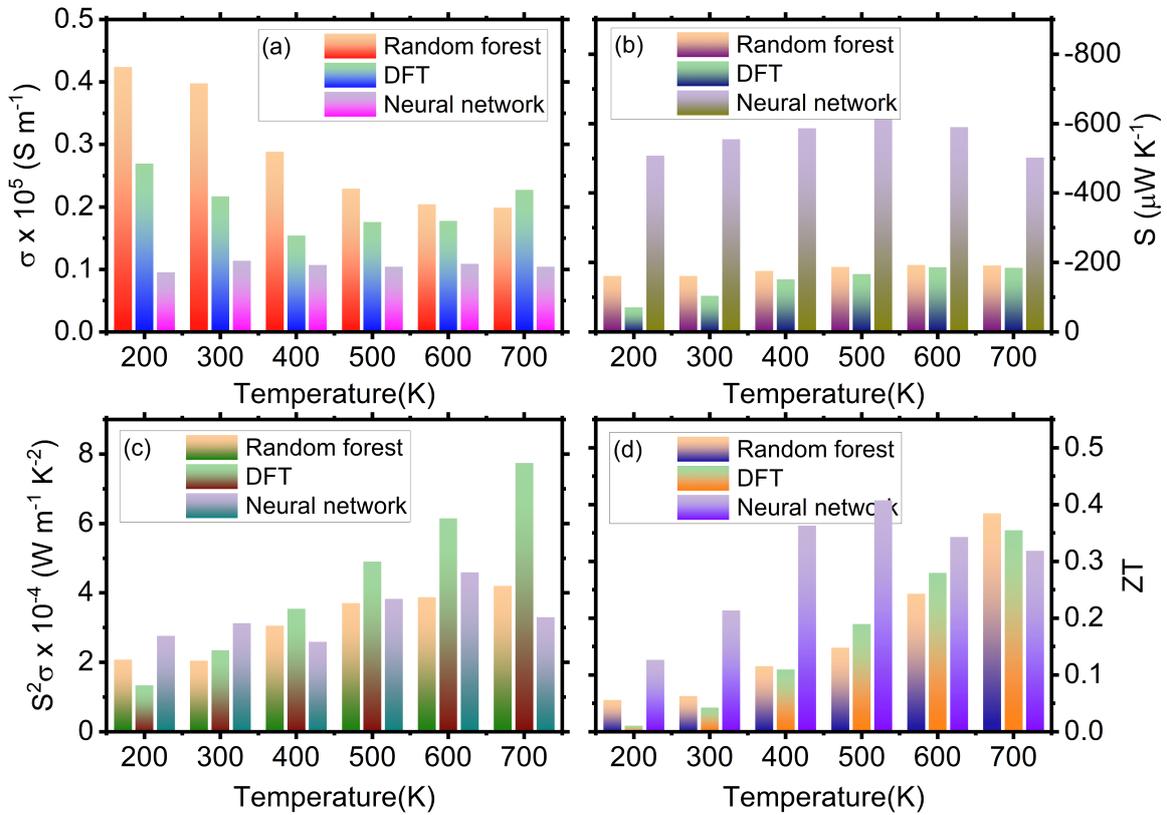

Figure 9 : Comparison of temperature-dependent (a) electrical conductivity, (b) Seebeck coefficient, (c) power factor, and (d) figure of merit of $Bi_2Se_3$ obtained from the machine learning and density functional theory.

**Table of content**

Hybrid density functional studies reveal the realization of topologically non-trivial surface states (TNSS) in $Bi_2Se_3$ by the application of uniaxial strain, which can be controlled and thereby control the thermoelectric parameters. Moreover, the integration of machine learning models, including random forest and neural network, into the study demonstrates their effectiveness in predicting temperature-dependent trends in thermoelectric properties.

Authors :

*Vipin K E, and Prahallad Padhan*

Title :

Advancements in Tuning Thermoelectric Properties: Insights from Hybrid Functional Studies, Strain Engineering, and Machine Learning Models

TOC Figure :

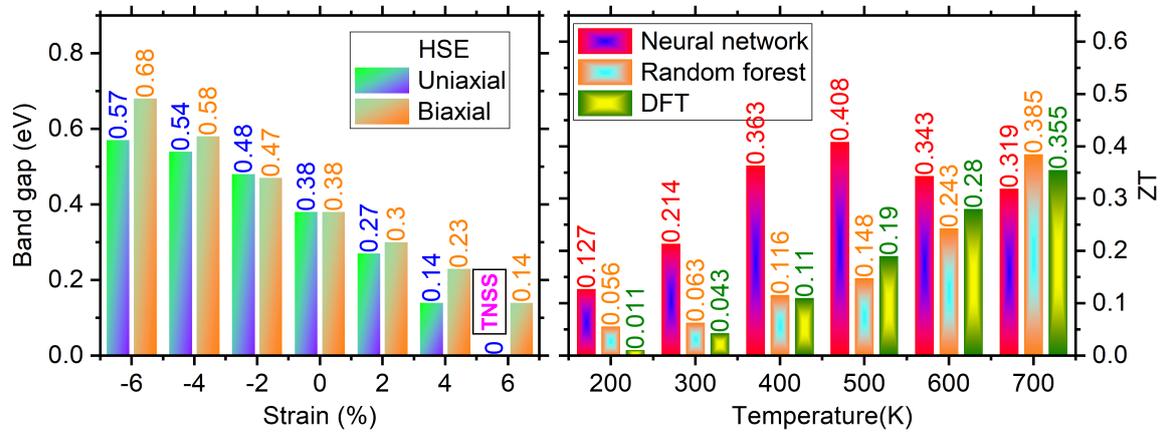